\documentclass{article}
\usepackage{graphicx}
\graphicspath{{converted_graphics/}}
\begin{document}

\begin{center}
{\LARGE Identification of a Novel \textquotedblleft
Fishbone\textquotedblright\ Structure}\vskip6pt

{\LARGE in the Dendritic Growth of Columnar Ice Crystals}\vskip6pt

{\Large Kenneth G. Libbrecht}\vskip4pt

{\large Department of Physics, California Institute of Technology}\vskip-1pt

{\large Pasadena, California 91125}\vskip-1pt

{\large address correspondence to: kgl@caltech.edu}\vskip-1pt

\vskip18pt

\hrule\vskip1pt \hrule\vskip14pt
\end{center}

\noindent \textbf{Abstract. Ice crystals growing in highly supersaturated
air at temperatures near -5 C exhibit a distinctive, nonplanar dendritic
morphology that has not been previously documented or explained. We examine
this structure and identify its most prominent features in relation to the
ice crystal lattice. Developing a full 3D numerical model that reproduces
this robust morphology will be an interesting challenge in understanding
diffusion-limited crystal growth in the presence of highly anisotropic
surface attachment kinetics.}

\vskip4pt \noindent\textit{[The figures in this paper have been reduced in
size to facilitate rapid downloading. The paper is available with higher
resolution figures at http://www.its.caltech.edu/\symbol{126}%
atomic/publist/kglpub.htm, or by contacting the author.]}

\section{Ice Dendrites}

Ice crystals growing in supersaturated air exhibit a remarkably rich variety
of morphological structures, including the many different types of snow
crystals that are commonly observed in the atmosphere \cite{fieldguide,
lowtemp, lowtemp2}. These different structures result from a complex
interplay of diffusion-limited growth and anisotropic attachment kinetics,
and much of the surface physics that is ultimately responsible for the
observed morphological diversity remains poorly understood \cite%
{libbrechtreview}.

At high water vapor supersaturations in air near one atmospheric pressure,
ice crystals tend to grow in complex dendritic structures. At temperatures
near -2 C or near -15 C, fern-like dendritic branching is the norm, and such
structures have been studied for several decades in the context of
diffusion-limited growth \cite{dendrites, libbrechtreview}. Ice crystal
dendrites at these temperatures are essentially two-dimensional, with the
entire branched structure confined mainly to a single plane. The growth
directions of the branches and sidebranches are also typically well aligned
along the a-axes of the ice crystal lattice.

\begin{figure}[p] % float placement: (h)ere, page (t)op, page (b)ottom, other (p)age
  \centering
  % file name: C:/Documents and Settings/Kenneth Libbrecht/My Documents/aatempfold/Fishbones/aaDSC_0215-arxiv.gif
  \includegraphics[bb=0 0 1270 824,width=5.67in,height=3.68in,keepaspectratio]{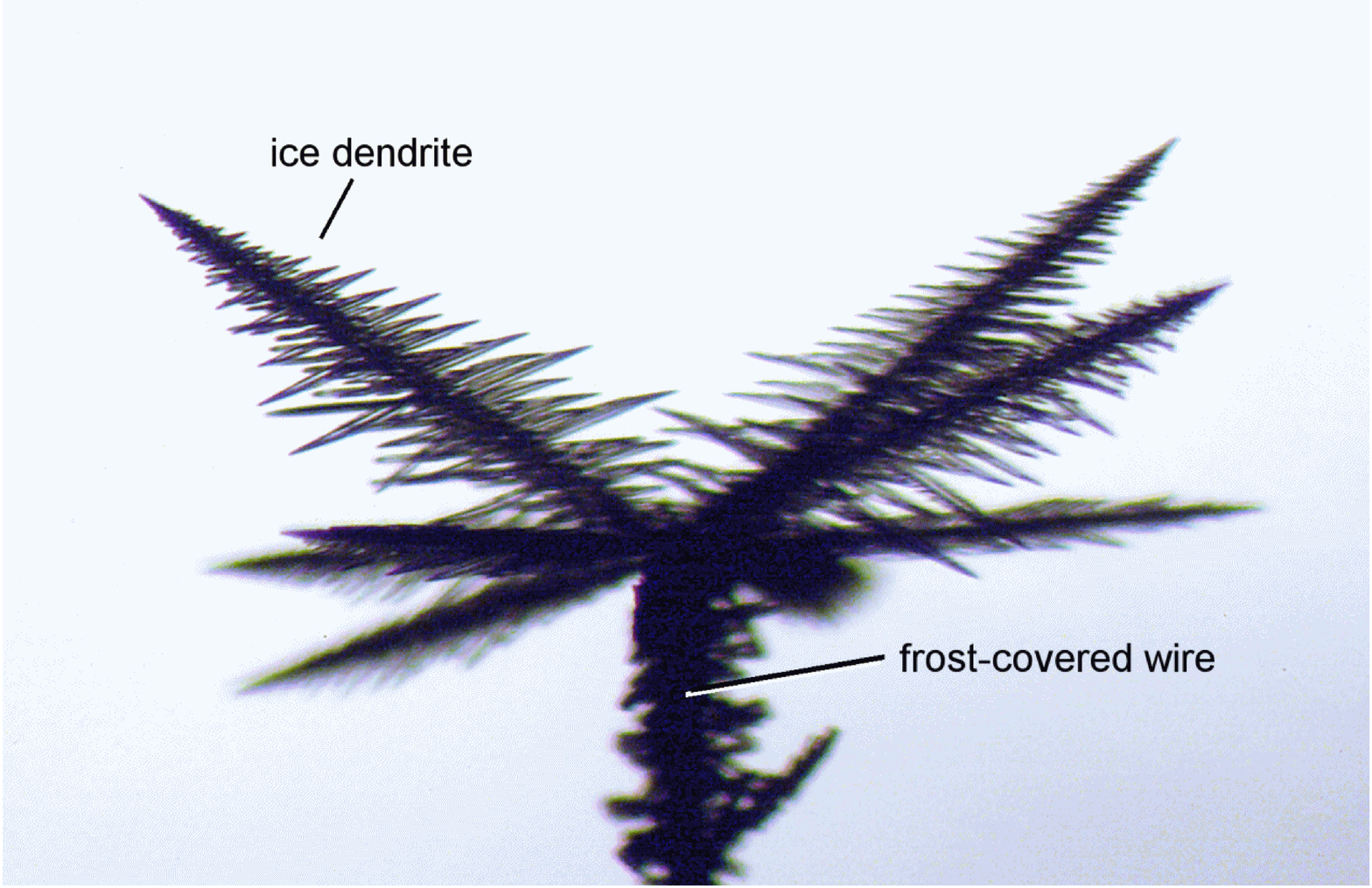}
  \caption{Ice dendrites growing near -5 C. These crystals were grown at a temperature near -5 C in a water vapor
diffusion chamber. The supersaturation was estimated to be roughly 100-200
percent in the absence of any growing crystals. This photograph shows
several ice dendrites, each about 2 mm in length, growing in various
directions from a frost-covered wire inside the chamber. The growth rates of
the dendrite tips were about 10-15 $\protect\mu $m/second \protect\cite%
{polar}. Each of the dendrites is a single ice crystal, with a fixed lattice
orientation throughout the main branch and the numerous sidebranches. The
eight dendrites seen in this image are separate crystals that nucleated at
different points on the wire and thus are oriented randomly with respect to
one another.}
  \label{dends}
\end{figure}

In contrast, ice dendrites growing in highly supersaturated air near -5 C
have a markedly different dendritic structure that includes nonplanar
branching along growth directions that are not aligned with respect to the
ice crystal lattice. Examples are shown in Figures \ref{dends} and \ref%
{single}. We observed these structures near -5 C whenever the
supersaturation was sufficiently high -- typically greater than 100-200
percent \cite{polar}. We have come to refer to them as \textquotedblleft
fishbones\textquotedblright\ for rather obvious reasons. These structures
are quite robust in that small changes in temperature or supersaturation do
not change the overall morphology, although the angles between the various
growth directions do depend on temperature and supersaturation. These
dendrites are quite easy to produce in water vapor diffusion chambers \cite%
{diffusion, diffusion2}, and it is likely they have been observed by a
number of researchers for several decades. To our knowledge, however, the
basic features of this crystal morphology have not been previously
identified.

\begin{figure}[tbp] % float placement: (h)ere, page (t)op, page (b)ottom, other (p)age
  \centering
  % file name: C:/Documents and Settings/Kenneth Libbrecht/My Documents/aatempfold/Fishbones/aaDSC_0355-arxiv.gif
  \includegraphics[bb=0 0 1082 871,width=5.67in,height=4.56in,keepaspectratio]{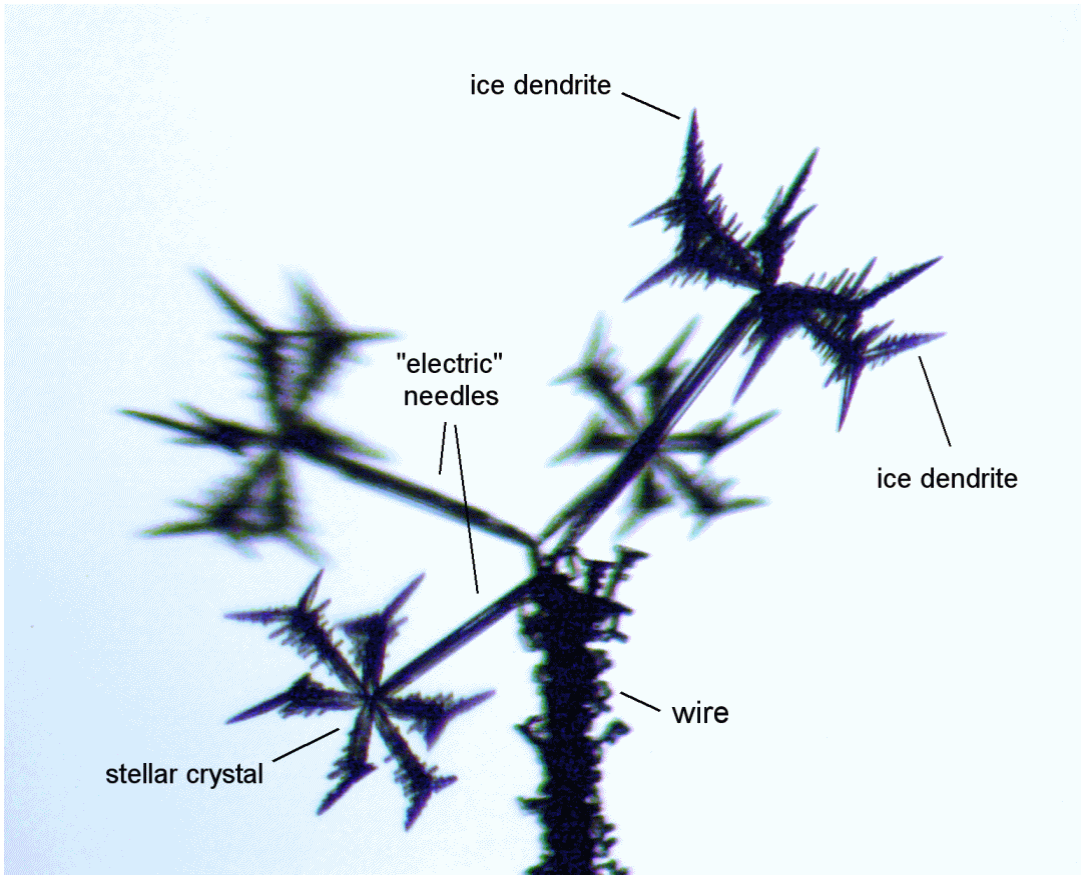}
  \caption{Defining the crystal axes. Beginning with a frost-covered wire, we
first grew \textquotedblleft electric\textquotedblright\ needle crystals as
described in \protect\cite{polar}. These needles were grown in the presence
of trace chemical impurities in the air, which produced needles growing
along the c-axis of the ice lattice \protect\cite{polar}. Once the needles
had grown out, we then removed the applied high voltage and moved the
assembly to a separate diffusion chamber, where growth at -15 C produced a
stellar plate crystal on the end of each needle. The supersaturation was
relatively low in this second chamber, so each stellar crystal had six
primary branches with little sidebranching. From there we moved the assembly
back to the first chamber, where fishbone dendritic structures at -5 C grew
from the tips of the stellar crystals. The level of chemical impurity was
quite low throughout the experiment, and the impurities seemed to have a
negligible affect on the normal growth of the fishbone dendrites, other than
perhaps shifting the growth angles slightly.}
  \label{electric}
\end{figure}

\begin{figure}[tbp] % float placement: (h)ere, page (t)op, page (b)ottom, other (p)age
  \centering
  % file name: C:/Documents and Settings/Kenneth Libbrecht/My Documents/aatempfold/Fishbones/aaDSC_0342-arxiv.gif
  \includegraphics[bb=0 0 900 760,width=5.67in,height=4.79in,keepaspectratio]{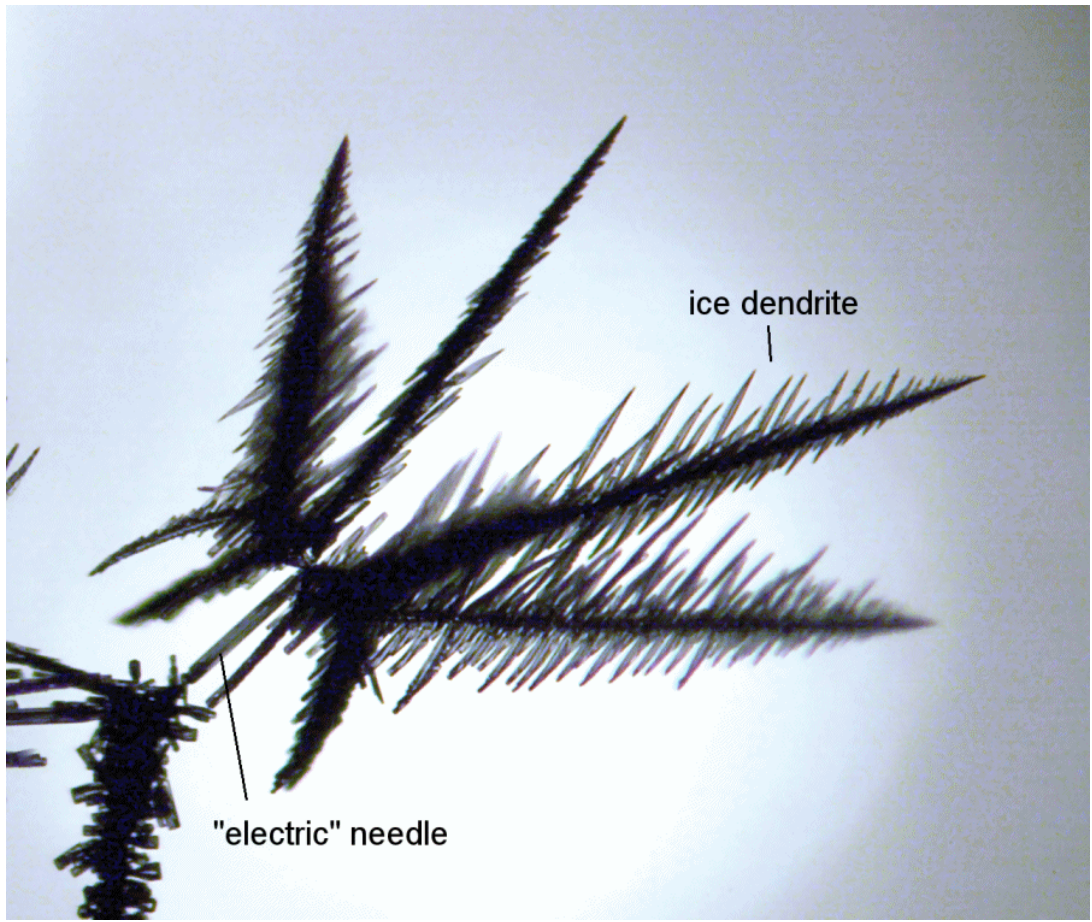}
  \caption{Similar to Figure 
\protect\ref{electric}, but with a closer view of the growth from a single
stellar crystal, after additional growth time. Fully developed fishbone
dendrites are growing out from each of the six branches of the stellar
crystal (although two dendrites are not clearly visible in this image). The
prograde and retrograde sidebranches are apparent on each dendrite.}
  \label{singleelectric}
\end{figure}

The relationship between the growth directions of the different dendritic
features and the underlying ice lattice was not immediately apparent, and we
found it useful to definitely establish the crystal axes in our experiments.
We did this by first growing \textquotedblleft electric\textquotedblright\
ice needles with stellar crystals on their ends, as described in \cite{polar}
and shown in Figures \ref{electric} and \ref{singleelectric}. Unfortunately,
still photographs do not easily convey the character of these complex
structures, and their three-dimensional morphology is best appreciated by
rotating a single specimen while viewing it with a stereo microscope.

\section{A Geometrical Model}

\begin{figure}[tbp] % float placement: (h)ere, page (t)op, page (b)ottom, other (p)age
  \centering
  % file name: C:/Documents and Settings/Kenneth Libbrecht/My Documents/aatempfold/Fishbones/aaVert1-arxiv.gif
  \includegraphics[bb=0 0 772 967,width=5.22in,height=6.54in,keepaspectratio]{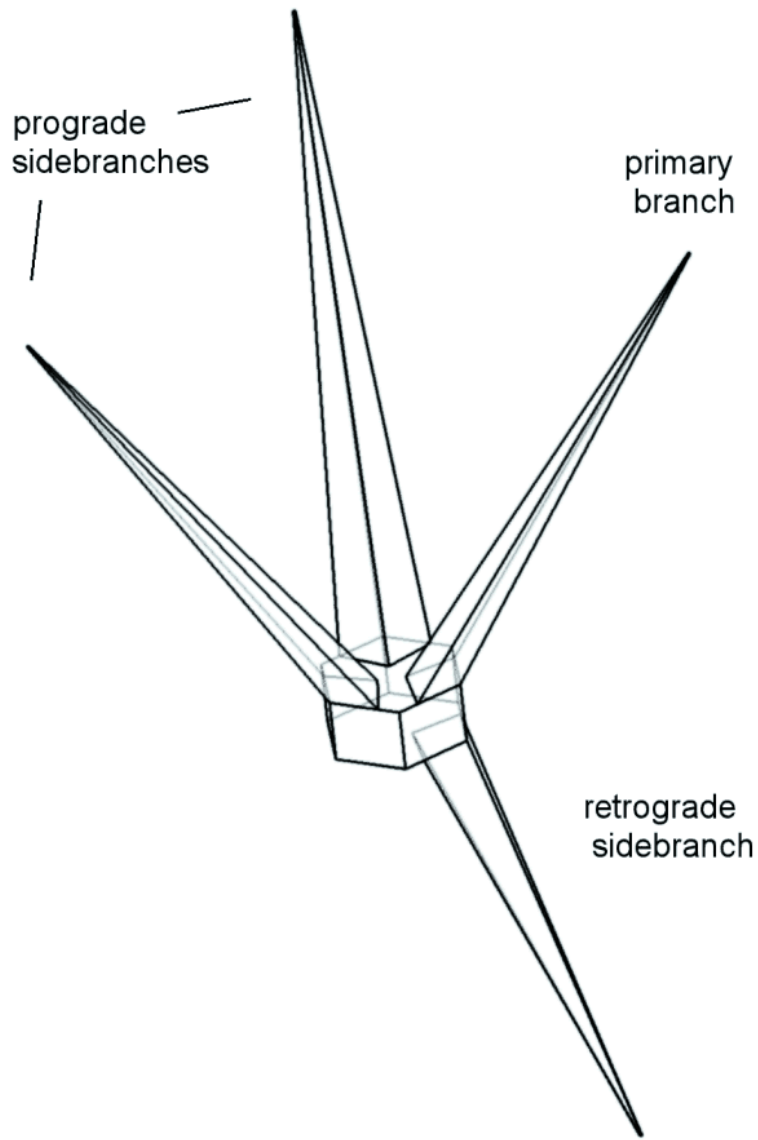}
  \caption{A single \textquotedblleft vertebra\textquotedblright\ in our
model of fishbone dendrites. The hexagonal prism is exaggerated and drawn
mainly to establish the crystal axes. Three \textquotedblleft
spires\textquotedblright\ grow from the top of the prism, along with a
single spire growing out from the bottom of the prism. Projected onto the
basal plane, each spire grows along a crystalline a-axis. The angles between
the spires and the c-axis are a function of the local temperature and
supersaturation. This vertebra includes three main free parameters -- the
angles between the c-axis and each of: the primary spire, the retrograde
spire, and the prograde spires.}
  \label{vert}
\end{figure}

\begin{figure}[tbp] % float placement: (h)ere, page (t)op, page (b)ottom, other (p)age
  \centering
  % file name: C:/Documents and Settings/Kenneth Libbrecht/My Documents/aatempfold/Fishbones/aa-twoviews-arxiv.gif
  \includegraphics[bb=0 0 1267 830,width=5.67in,height=3.71in,keepaspectratio]{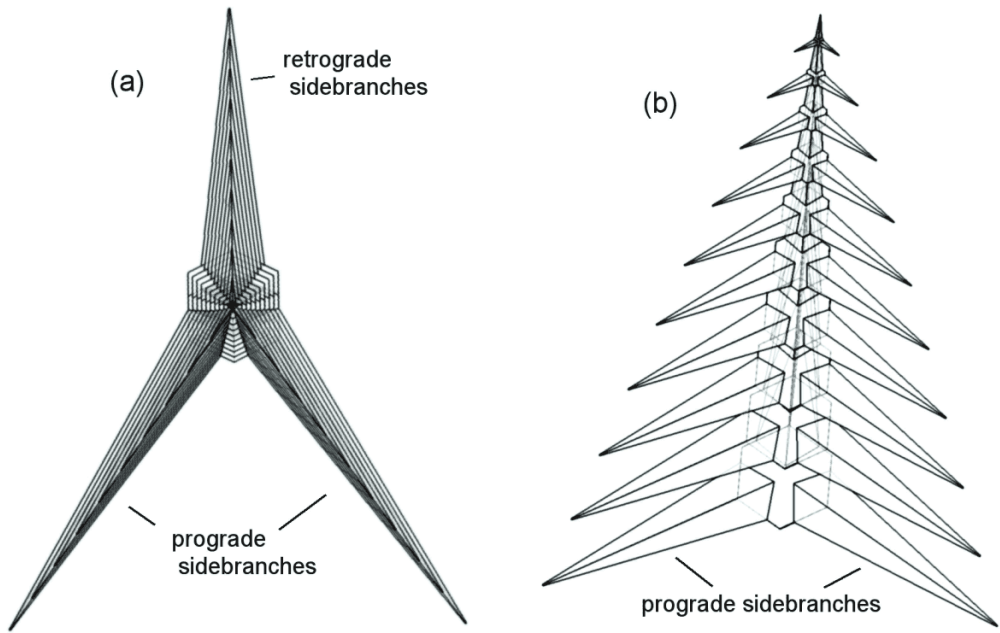}
  \caption{(a) The view straight down the primary branch. The angle
subtended by the two sets of prograde sidebranches depends on growth
conditions. Near -5 C, we found that this angle increases with increasing
supersaturation. (b) The view down the c-axis. From this viewing direction
the projected angle between the two sets of prograde sidebranches is 120
degrees, as are the angles between the prograde sidebranches and the primary
branch. Regardless of the particular growth directions of the different
branches and sidebranches (which depend on temperature and supersaturation),
it is always possible to find a viewing angle that exhibits this 120-degree
symmetry.}
  \label{twoviews}
\end{figure}

Once we established the crystal axes with certainty and viewed a number of
growing dendrites under different conditions, we were able to create a
geometrical model of their structure, as shown in Figures \ref{vert}, \ref%
{assembly}, and \ref{twoviews}. This model nicely fits all our observations
and shows how the three-dimensional dendritic structure arises from the
intrinsic hexagonal symmetry of the ice crystal lattice.

It seems possible that our single-crystal fishbone dendrites could be
misidentified as the polycrystalline \textquotedblleft
spearhead\textquotedblright\ structures described in \cite{lowtemp}. The
distinctive fishbone sidebranching would be less apparent at lower
supersaturations, giving the two forms a quite similar appearance. Further
investigation of the fishbone morphology as a function of temperature and
supersaturation, including observations of the transition from columnar to
plate-like dendrites, would elucidate this further.

At present we have no clear constraints on the various growth angles in our
geometrical model, as these depend on growth conditions and are tied to the
highly anisotropic surface attachment kinetics inherent in ice crystal
growth. Numerical techniques using cellular automata are capable of modeling
the diffusion-limited growth of realistic faceted dendritic structures in
three dimensions \cite{gg3d}, but the required input physics remains quite
uncertain \cite{libbrechtmodel}.

We believe it will be an interesting challenge to create a numerical growth
model that reproduces the morphology of fishbone dendrites. On one hand,
this overall morphology is robust with respect to rather large changes in
growth conditions (which is also true for -15 C dendrites), so we expect
that models would reproduce the overall fishbone features even without a
perfect parameterization of the input physics. On the other hand, the
fishbone structure is sufficiently complex that significant advances in our
physical understanding of the numerical models may be required. In any case,
comparing theoretical models with quantitative measurements of growing
dendrites will likely shed light on the presently enigmatic molecular
dynamics that governs ice crystal growth.

\begin{figure}[tbp] % float placement: (h)ere, page (t)op, page (b)ottom, other (p)age
  \centering
  % file name: C:/Documents and Settings/Kenneth Libbrecht/My Documents/aatempfold/Fishbones/aaDSC_0310-arxiv.gif
  \includegraphics[bb=0 0 888 918,width=5.67in,height=5.86in,keepaspectratio]{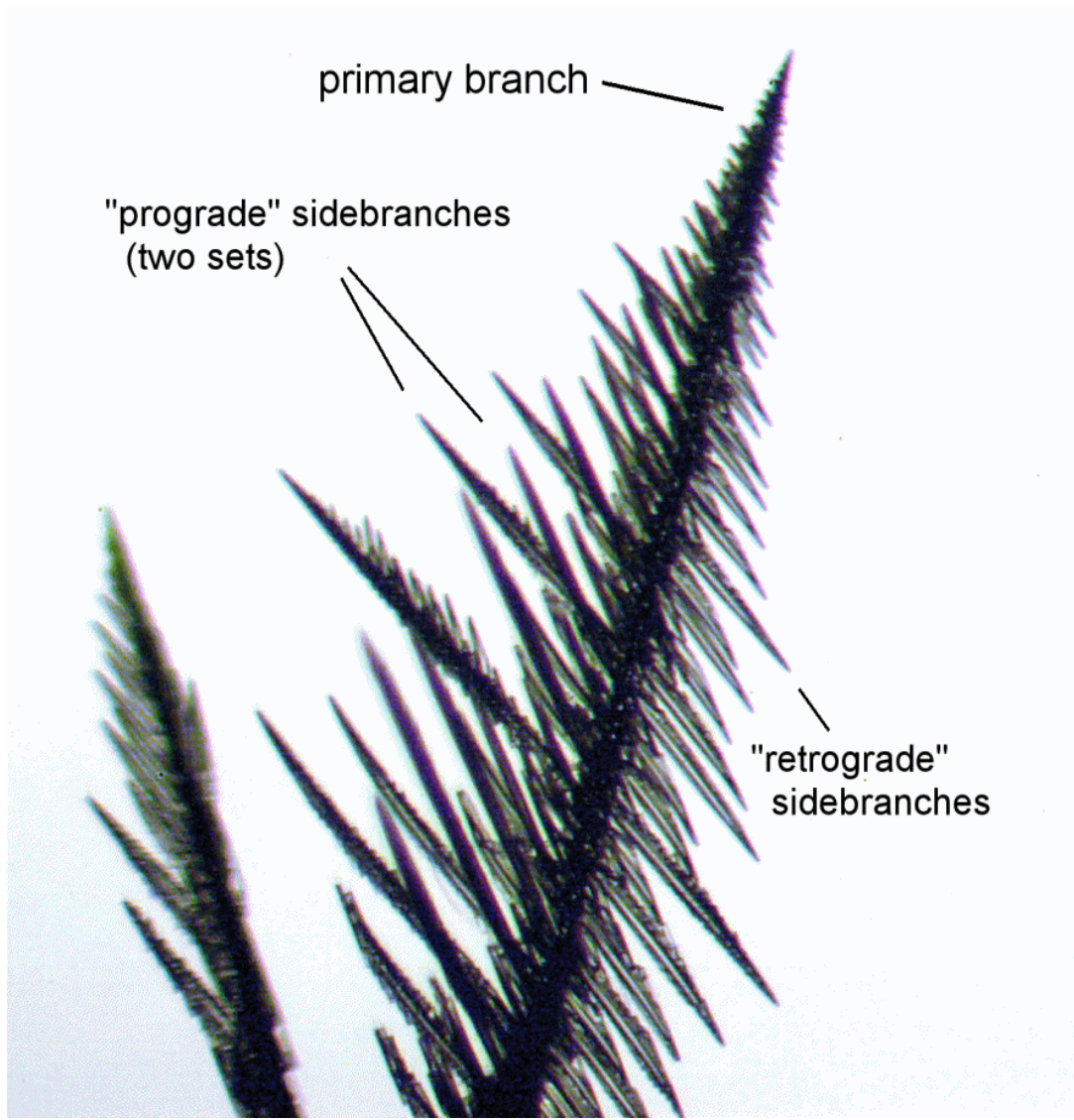}
  \caption{A closer view of a single ice
dendrite growing near -5 C. As is typical for the diffusion-limited growth
of dendritic structures \protect\cite{dendrites}, the primary branch grows
outward with a nearly constant tip velocity. For growth from vapor, the tip
velocity is approximately proportional to supersaturation \protect\cite%
{polar}. Two sets of \textquotedblleft prograde\textquotedblright\
sidebranches grow out from the primary spine, along with one set of
\textquotedblleft retrograde\textquotedblright\ sidebranches. The angle
between the growth directions of the prograde sidebranches and that of the
primary spine is less than 90 degrees; the angle between the retrograde
sidebranches and the primary spine is greater than 90 degrees. The two sets
of prograde sidebranches lie in two separate planes, and the angle between
the planes is typically less than 60 degrees. The growth directions of the
different features depend on both the temperature and supersaturation inside
the growth chamber. These dendritic structures are quite robust in that they
are the normal growth forms near -5 C when the supersaturation is high.}
  \label{single}
\end{figure}

\begin{figure}[tbp] % float placement: (h)ere, page (t)op, page (b)ottom, other (p)age
  \centering
  % file name: C:/Documents and Settings/Kenneth Libbrecht/My Documents/aatempfold/Fishbones/aa-assembly1-arxiv.gif
  \includegraphics[bb=0 0 1090 1650,width=4.7in,keepaspectratio]{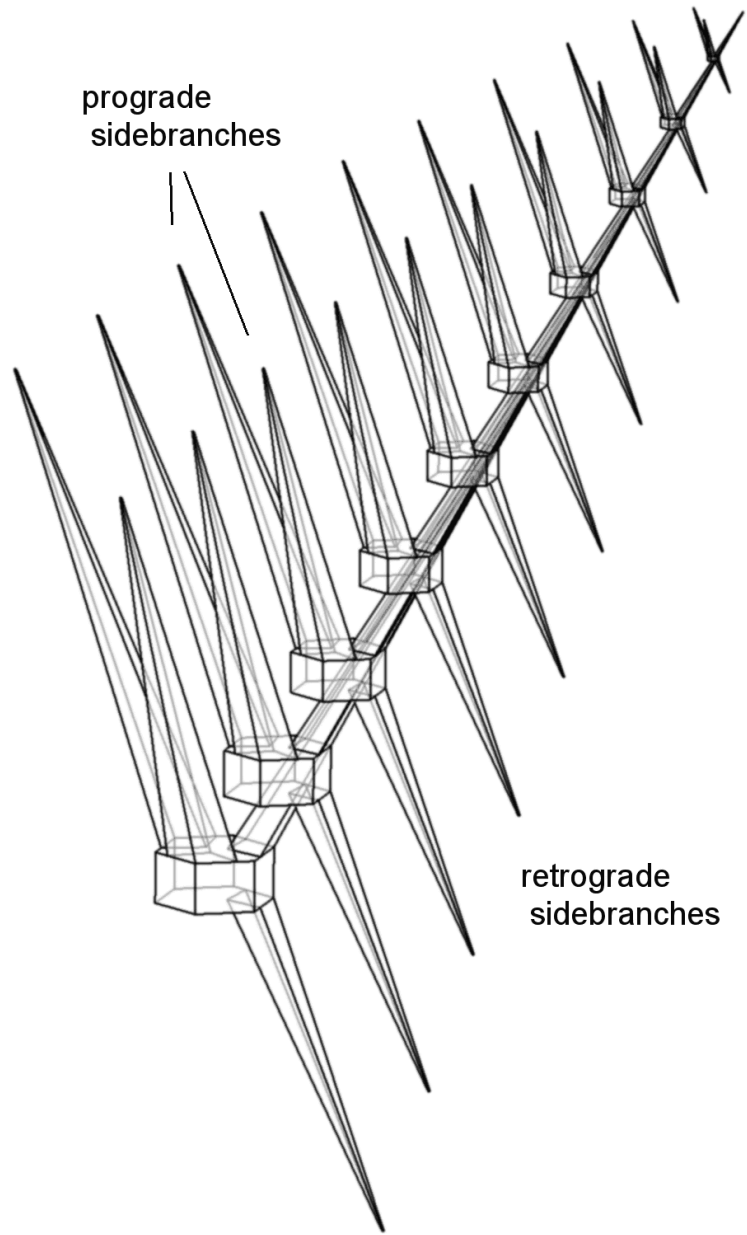}
  \caption{Assembly of several
\textquotedblleft vertebrae\textquotedblright\ into a growing fishbone
dendrite. Again the hexagonal prisms are drawn mainly to indicate the
crystal axes; they are not meant to be a visible part of the final
structure. This geometric model should be compared with the actual ice
dendrite shown in Figure \protect\ref{single}.}
  \label{assembly}
\end{figure}

\end{document}